# A 33-year NPP monitoring study in southwest China by the fusion of multi-source remote sensing and station data

Xiaobin Guan [1], Huanfeng Shen [1,2,*], Wenxia Gan [3], Gang Yang [4], Lunche Wang [5], Xinghua Li [6] and Liangpei Zhang [2,7]

[1] School of Resource and Environmental Sciences, Wuhan University, Wuhan 430079, Hubei, China; guanxb@whu.edu.cn

[2] Collaborative Innovation Center of Geospatial Technology, Wuhan 430079, Hubei, China; shenhf@whu.edu.cn

[3] School of Resource and Civil Engineering, Wuhan Institute of Technology, Wuhan 430205, Hubei, China;

[4] Department of Geography and Spatial Information Techniques, Ningbo University, Ningbo 315211, Zhejiang, China;

[5] Laboratory of Critical Zone Evolution, School of Earth Sciences, China University of Geosciences, Wuhan 430074, Hubei, China;

[6] School of Remote Sensing and Information Engineering, Wuhan University, Wuhan 430079, Hubei, China;

[7] The State Key Laboratory of Information Engineering in Surveying, Mapping and Remote Sensing, Wuhan University, Wuhan 430079, Hubei, China;

**\*** Correspondence: shenhf@whu.edu.cn (H.S)

**Abstract:** Knowledge of regional net primary productivity (NPP) is important for the systematic understanding of the global carbon cycle. In this study, multi-source data were employed to conduct a 33-year regional NPP study in southwest China, at a 1-km scale. A multi-sensor fusion framework was applied to obtain a new normalized difference vegetation index (NDVI) time series from 1982 to 2014, combining the respective advantages of the different remote sensing datasets. As another key parameter for NPP modeling, the total solar radiation was calculated by the improved Yang hybrid model (YHM), using meteorological station data. The verification described in this paper proved the feasibility of all the applied data processes, and a greatly improved accuracy was obtained for the NPP calculated with the final processed NDVI. The spatio-temporal analysis results indicated that 68.07% of the study area showed an increasing NPP trend over the past three decades. Significant heterogeneity was found in the correlation between NPP and precipitation at a monthly scale, specifically, the negative correlation in the growing season and the positive correlation in the dry season. The lagged positive correlation in the growing season and no lag in the dry season indicated the important impact of precipitation on NPP. What is more, we also confirmed that the variation of NPP was driven by different factors in three distinct stages. Significant climate warming led to a great increase of NPP from 1992 to 2002, while NPP clearly decreased during 1982–1992 and 2002–2014 due to the frequent droughts caused by the precipitation decrease.

## 1. Introduction

As a key component of the global carbon cycle, the terrestrial ecosystem is the main force that can uptake free carbon from the atmosphere and convert it into organic compounds [1,2]. An improved understanding of the terrestrial ecosystem carbon cycle is urgently needed under the background of intense global climate change [3-5]. Net primary productivity (NPP), which is the residual amount of organic matter produced by vegetation photosynthesis minus its autotrophic respiration consumption, is an important ecological indicator for the status of a terrestrial ecosystem carbon budget [6,7]. NPP can be precisely acquired by field measurements at a site level, while model estimation is an efficient approach for regional or larger scales. A large number of models have been proposed in previous studies [8], and with the development of the remote sensing, satellite data based models have been extensively applied to study terrestrial NPP. The reason for the popularity of the satellite data based models is that remote sensing images can provide continuous, dynamic, and comprehensive land-surface information for any region around the earth [9,10]. Satellite land-cover data and spectral vegetation index products (i.e., the normalized difference vegetation index, NDVI) are the most commonly used core data when modeling NPP for a large region [8].

Numerous studies have explored the spatio-temporal patterns of NPP based on moderate-resolution satellite NDVI datasets [11-13], such as the Advanced Very High Resolution Radiometer (AVHRR) data from the National Oceanic and Atmospheric Administration (NOAA) satellites and the Moderate Resolution

Imaging Spectroradiometer (MODIS) data from the Terra/Aqua satellites [14,15]. However, the related studies have mostly concentrated on the variation of NPP at global or continental scales [12,16-18]. Although some research has concerned the regional carbon cycle [19-21], the in-depth studies are still lacking. Nevertheless, regional terrestrial ecosystem carbon budgets are important, because the characteristics of NPP can highly vary in space and time due to the different natural environments and human landscapes [22,23]. Further study of regional carbon cycles is therefore necessary for the systematic understanding of the global carbon cycle, since an integrated and dense carbon observation and analysis system is urgently required [24].

In regional carbon cycle studies with satellite data based models, the remote sensing data quantity and quality are the determinants for a significant and in-depth analysis. MODIS datasets have been applied in many studies of the spatio-temporal variations of regional NPP and its potential causal factors, at spatial resolutions of 250 m to 1 km [23,25-27]. Although these studies have captured the spatial information of NPP, they have been unable to analyze the NPP variation and its relationship with environmental factors at a long-term level, because only data covering the last decade are available, and no data are available before the year 2000. The same limitation also exists in the studies with Satellite Pour l'Observation de la Terre (SPOT) Vegetation products [28-30];. Studies using Landsat data can offer even more detailed spatial information, but it is difficult to acquire a continuous long-telrm series, because of the poor temporal resolution and the influence of cloud cover [31-33]. Since the 1980s, AVHRR datasets have been extensively employed to study the regional NPP of a long time period [19,34-36], but they have been generally used for much larger regions, due to the coarse spatial resolution of 8 km. Furthermore, it has been proven in many papers that a coarse resolution can lead to an obvious accuracy loss when modeling NPP, as a result of the spatial heterogeneity of the data [37-39]. In general, further study of the regional carbon cycle is limited by the inter-inhibitive characteristics of the different sensors. Therefore, the key challenge when using remote sensing measurements in regional NPP studies is to achieve a level of long-term consistency and accuracy [24]. Integrating remote sensing data from different sensors and synthesizing their respective merits is the only way to settle the problem [40-42].

Yunnan province, which is the most southwestern part of China, is one of the most important carbon sinks in the continent [43]. However, the region has suffered from more and more frequent droughts in recent decades [44,45]. A continuous four-year extreme drought occurred from 2009 to 2012, which had a disastrous impact [46]. Although a number of studies have investigated the drought impacts in recent years [46,47], meritorious research into the long-term variation of the regional carbon cycle and its relationship with climatic factors is extremely rare. As a result, to address these issues, the main objectives of this paper are: 1) to develop an innovative framework for the generation of a continuous 33-year NPP time series for Yunnan province at a 1-km spatial resolution with the Carnegie-Ames-Stanford Approach (CASA) model by fusing multi-source remote sensing data and station data; and 2) to carry out an in-depth analysis of the spatial and temporal characteristics of the terrestrial ecosystem carbon cycle in the study area, as well as its relationships with the climate.

## 2. Data sources

### 2.1. Study area

Yunnan province, which is located in southwest China between 21.13°–29.25°N and 97.52°–106.18°E, was chosen as the study area in this research (Figure 1). Yunnan covers a total area of 394,000 $km^2$, and mountainous landforms occupy more than 90% of the region. The unique mountainous landforms lead to a diverse climate and provide conditions for the growth of many different vegetation species. Vegetation covers most of the area (approximately 94%), and the rich and varied vegetation types range from tropical species to frigid species [48]. Most of the territory locates in the subtropical or tropical zones, with the Tropic of Cancer running through its southern part. The synergistic effects of the tropical/subtropical climate, the monsoon climate, and the mountain climate lead to highly complex climate patterns in the area. Although the temperature is moderate and rainfall is abundant, the uneven intra-annual allocation usually results in wet summers and dry winters [46]. The complex terrain and climate make it even more important to study the carbon cycle in such an important carbon sink, particularly the impacts of climate change and the frequent droughts.

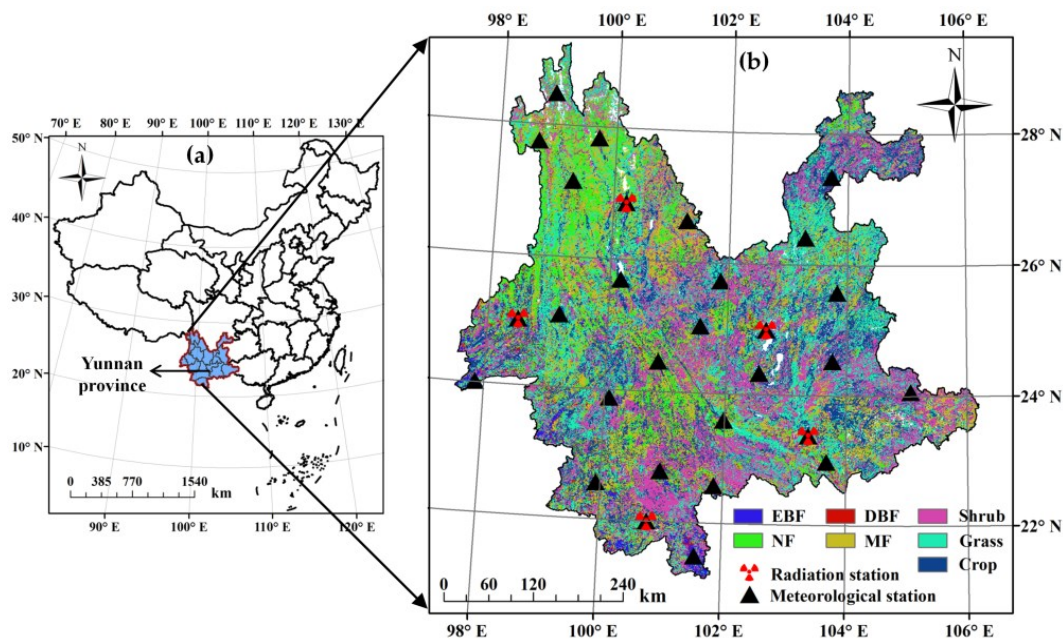

**Figure 1.** Study area: (**a**) location of Yunnan province in China; (**b**) spatial distribution of vegetation cover and radiation/meteorological stations (EBF: evergreen broadleaf forest; DBF: deciduous broadleaf forest; NF: needle-leaf forest; MF: mixed forest).

*2.2. Data sources*

### 2.2.1. NDVI datasets

The Global Inventory Modelling and Mapping Studies 3rd generation (GIMMS3g) NDVI product from the NOAA/AVHRR and MODIS monthly NDVI collection (MOD13A3) were selected as the basis to composite the 33-year NDVI time series at a 1-km scale. The GIMMS3g dataset from 1982 to 2012 was obtained from the National Aeronautics and Space Administration (NASA) Ames Ecological Forecasting Lab. The spatial resolution of this dataset is 8 km, and the time interval is half a month. The dataset has been proven to have a better availability and quality than other AVHRR-based NDVI products [49]. The MOD13A3 collection from 2000 to 2014 (no data are available for January 2000) was acquired from the NASA Earth Observing System (EOS) program, at a spatial resolution of 1 km. The data, which are obtained based on the spectral bands that are primarily designed for the study of vegetation and the land surface [50], have been widely applied in numerous vegetation studies [13]. In this study, the maximum value compositing (MVC) technique was employed to obtain the monthly GIMMS3g NDVI, to match the time interval of the MODIS data.

### 2.2.2. Meteorological datasets

The meteorological datasets over the study period of 1982–2014 were obtained from the China Meteorological Administration (CMA), including monthly and daily precipitation and air temperature, and daily surface pressure, air relative humidity, and sunshine duration from the 29 uniformly distributed meteorological stations (Figure 1 (b)). In addition, radiation datasets of the five specified radiation stations (Figure 1 (b)) were also compiled. The station records were carefully interpolated into the same spatial resolution as the MODIS NDVI (1 km×1 km), using the Australian National University SPLINe (ANUSPLIN) package [51], with elevation or slope data as the independent covariates. This package can be used to undertake professional meteorological interpolation using the thin plate smoothing splines surface-fitting technique. The use of independent covariates can further improve the precision. After repeated attempts, it was concluded that elevation was the optimal covariate for temperature, and slope was the optimal covariate for precipitation and radiation.

### 2.2.3. Other data

The WESTDC2.0 land-cover map was derived by the Chinese Academy of Sciences (CAS) Environmental and Ecological Science Data Center for West China (WESTDC) [52]. The map takes full advantage of the 1:100,000 land resources data surveyed by CAS, and is integrated with multi-source satellite classification information. In this study, the data were synthesized into eight classes: evergreen broadleaf forest (EBF), deciduous broadleaf forest (DBF), needle-leaf forest (NF), mixed forest (MF), shrub, grass, crop, and other land covers.

The measurement-based biomass/NPP datasets from Luo's study [53], which have been used in many studies [47,54], were employed as the validation data in this study. All the records over the study area were from the Yunnan Ministry of Forestry for the year of 1983. The data include the forest biomass/NPP for most of the plant components, and the location and dominant species of each site. As the records of NPP were provided with the unit of dry matter (t DM ha$^{-1}$ year$^{-1}$), a conversion factor of 50 was needed to change this into carbon content (gC m$^{-2}$ year$^{-1}$) [54,55].

## 3. Method

### 3.1. CASA model

The Carnegie-Ames-Stanford Approach (CASA) model, which was developed on the basis of light-use efficiency, served to estimate the monthly NPP in the study area [56,57]. The calculation of NPP can be expressed as the product of absorbed photosynthetic active radiation ($APAR$, MJ m$^{-2}$) and the light-use efficiency ($\varepsilon$, gC MJ$^{-1}$), as follows:

$$NPP(x,t) = APAR(x,t) \times \varepsilon(x,t), \tag{1}$$

where $NPP(x,t)$ is the total fixed NPP of pixel $x$ in month $t$, $APAR(x,t)$ is the total amount of absorbed photosynthetic active radiation over the period, and $\varepsilon(x,t)$ is the actual light-use efficiency. The calculation of $APAR$ and $\varepsilon$ is shown below:

$$APAR(x,t) = R_s(x,t) \times 0.5 \times FPAR(x,t), \tag{2}$$

$$\varepsilon(x,t) = \varepsilon^*(x,t) \times T_1(x,t) \times T_2(x,t) \times W(x,t), \tag{3}$$

where $R_s(x,t)$ is the total solar radiation of pixel $x$ in month $t$; the coefficient 0.5 is the approximate ratio of photosynthetic active radiation (0.4–0.7 μm) to total solar radiation; and $FPAR(x,t)$ is the fraction of photosynthetic active radiation absorbed by the vegetation canopy, which is determined by the NDVI and vegetation type.

$$FPAR(x,t) = \min[(SR(x,t) - SR_{\min}) / (SR_{\max} - SR_{\min}), 0.95], \tag{4}$$

$$SR(x,t) = [1 + NDVI(x,t)] / [1 - NDVI(x,t)], \tag{5}$$

where $SR(x,t)$ is the simple ratio of NDVI; $SR_{\max}$ and $SR_{\min}$ are the constants related to the vegetation type, which is shown in Table 1 [58].

**Table 1.** The value of maximum light utilization efficiency ($\varepsilon^*$), $SR_{\min}$, $SR_{\max}$ for different vegetation types.

| Vegetation type | EBF | DBF | NF | MF | Shrub | Grass | Crop |
|---|---|---|---|---|---|---|---|
| $\varepsilon^*$ (gC MJ$^{-1}$) | 0.985 | 0.692 | 0.485 | 0.768 | 0.429 | 0.542 | 0.542 |
| $SR_{\min}$ | 1.050 | 1.050 | 1.050 | 1.050 | 1.050 | 1.050 | 1.050 |
| $SR_{\max}$ | 5.170 | 6.910 | 6.630 | 4.670 | 4.490 | 4.460 | 4.460 |

Note: $\varepsilon^*$ is the maximum light-use efficiency; $SR_{min}$ is the factor $SR$ for unvegetated land areas; $SR_{max}$ approximates the values of $SR$ when all solar radiation is intercepted; EBF: evergreen broadleaf forest; DBF: deciduous broadleaf forest; NF: needle-leaf forest; MF: mixed forest).

In Eq. (3), $\varepsilon^*(x,t)$ is the maximum light-use efficiency, the value of which varies with the vegetation type according to the previous study of ecosystems in China (Table 1) [58]; $T_1(x,t)$ and $T_2(x,t)$ are the temperature stress factors; and $W(x,t)$ is the moisture stress factor.

$$T_1(x,\ t)=0.8+0.02\times T_{opt}(x)-0.0005\times T_{opt}(x)\times T_{opt}(x),\tag{6}$$

$$T_2(x,\ t)=1.1814\ /\ \{1+e^{[0.2(T_{opt}(x)-10-T(x,t))]}\}\ /\ \{1+e^{[0.3(-T_{opt}(x)-10+T(x,t))]}\},\tag{7}$$

$$W(x,t)=0.5+0.5EET(x,t)\ /\ PET(x,t),\tag{8}$$

where $T_{opt}(x)$ is the temperature when NDVI reaches its maximum for the year; $T(x,t)$ is the monthly mean temperature; $EET(x,t)$ and $PET(x,t)$ are the soil properties derived from the sub-model of regional evapotranspiration [59].

The spatial and temporal resolution of the NPP from the CASA model is determined by the resolution of the NDVI. In order to obtain a long-term NDVI time series with a suitable resolution, image processing algorithms were applied to integrate the respective advantages of the MODIS and GIMMS3g datasets. In addition, the total solar radiation was precisely calculated by the improved YHM model. The overall workflow of NPP estimation is depicted in Figure 2.

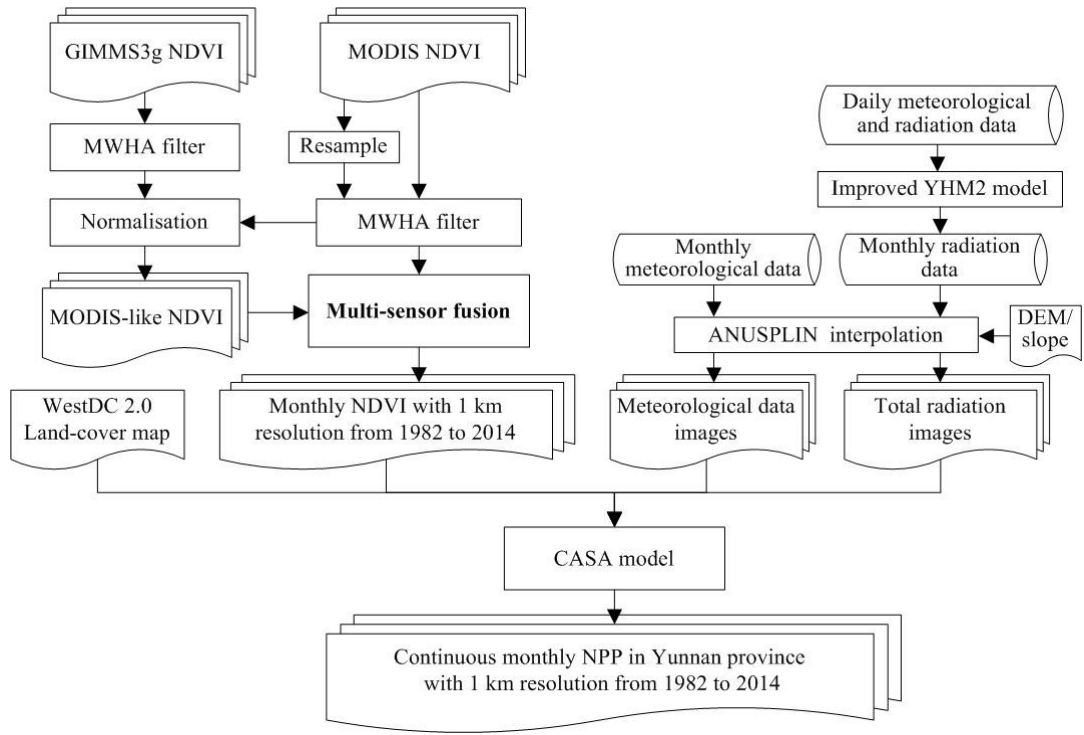

**Figure 2.** The workflow of NPP estimation.

### 3.2. Generation of the 33-year NDVI time series at a 1-km scale

The three processes described below were adopted to eliminate the problems with regard to data quality, sensor differences, and the coarse spatial resolution, for the NDVI datasets.

### 3.2.1. NDVI filtering

The application of NDVI time-series data is usually limited by the existence of unwanted noise and errors caused by the cloud presence and other atmospheric contamination [60]. In order to obtain a high-quality NDVI time series, the moving weighted harmonic analysis (MWHA) [61] method was employed to correct these contaminated values in the GIMMS3g and MODIS NDVI datasets. The algorithm was proposed based on a modification of the harmonic analysis method, which has been proved to be a better strategy for different NDVI datasets with various time intervals, and is reliable for pixels in the vegetation dormancy period [61]. The filter result of an example pixel for the period from 2007 to 2008 is shown in Figure 3, where it can be observed that the sudden abnormal drops in the two NDVI time series have all been exactly adjusted. After the filtering, the two time series show much more similar variation curves and approach the actual vegetation variation. After sampling several pixels, we concluded that the data quality of the MODIS dataset was obviously better than the GIMMS3g time series, with far fewer contaminated points.

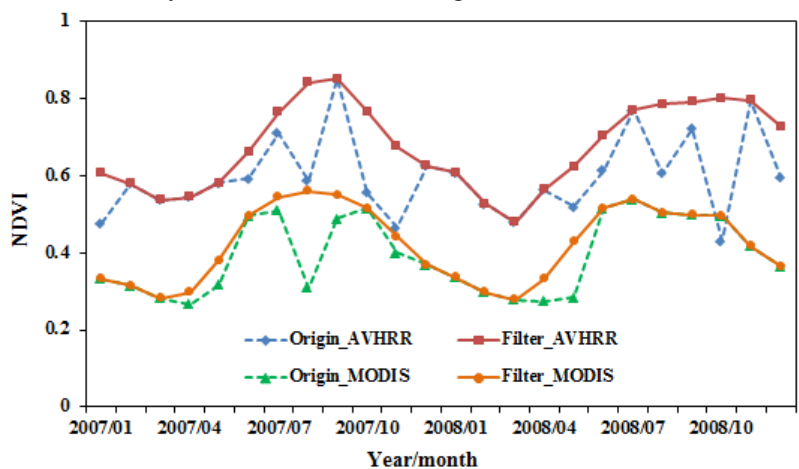

**Figure 3.** Filter results of the MODIS and AVHRR NDVI time series by the MWHA method for an example of a pixel located at (25.32N, 103.53E), from 2007 to 2008.

### 3.2.2. Normalization

A large gap between the NDVI of the two sensors for the specific pixel can be observed in Figure 3, and the NDVI values of MODIS are generally lower than GIMMS3g, with a mean difference of 0.07 in the study area. This is likely to be caused by the disparity in the characteristics of the two sensors, and the fact that no atmospheric correction has been applied to the GIMMS3g data [62,63]. Fensholt and Proud concluded that the temporal trends derived from the GIMMS NDVI agree well with the MODIS data [64], overall, so a unary linear regression normalization model can be used to express the relationship between the two datasets. In this study, the GIMMS3g data were normalized to be the same as MODIS using a pixel-by-pixel linear regression method, at the scale of the GIMMS3g data (8 km). For each pixel, a linear relationship was obtained based on the data in the mutual time period from February 2000 to December 2012 (155 pairwise images). Then it was applied to the GIMMS3g data for the corresponding pixels before the year 2000. Thus, the data time series from 1982 to 1999 can be considered as MODIS-like NDVI, with data values consistent with MODIS, but with the same resolution as GIMMS3g.

### 3.2.3. Multi-sensor fusion

Although the sensor differences were removed, gaps still existed in the spatial resolution between the MODIS NDVI and the obtained MODIS-like time series. On account of the 8-km resolution being too coarse (Figure 5 (a)), which can lead to overestimation when modeling NPP [37,38], multi-sensor fusion was an effective way to improve the spatial resolution of the MODIS-like data [65-67]. Many vegetation studies have applied the spatial and temporal adaptive reflectance fusion model (STARFM) and the extended STARFM (ESTARFM) to vegetation index fusion and prediction [40,68-71]. In this study, a spatio-temporal information fusion method based on a non-local means filter was employed to improve the spatial resolution of the MODIS-like data [72].

Multi-sensor fusion can be used to predict the MODIS NDVI value at $t_1$ based on the MODIS-like data at $t_1$ and the reference MODIS and MODIS-like images acquired at $t_0$. The prediction of the fine-resolution NDVI before the year 2000 can be expressed as:

$$F\left(x_{p/2}, y_{p/2}, t_1\right) = \sum_{i=1}^{p} \omega_i \times \left(F\left(x_i, y_i, t_0\right) + C\left(x_i, y_i, t_1\right) - C\left(x_i, y_i, t_0\right)\right), \tag{9}$$

where $F$ and $C$ represent the fine-resolution and coarse-resolution NDVI, respectively; $t_0$ is the acquisition date of the reference data; $t_1$ is the prediction date; $\left(x_{p/2}, y_{p/2}\right)$ is the location of the predicted pixel; $\left(x_i, y_i\right)$ denote the pixel location; $p$ is the size of the moving window; and $\omega_i$ is the spatial weighting function. The innovation of the applied fusion algorithm is the more reasonable calculation of $\omega_i$, which takes full consideration of the spatial relationship between pixels based on the concept of the non-local means filter. Considering the seasonal inconsistency of NDVI patterns, the fine-resolution NDVI of each month from 1982 to 1999 was predicted referring to the MODIS data for the corresponding month in the nearest year.

### 3.3. Accurate calculation of total solar radiation with the improved YHM model

The surface total solar radiation ($R_s$) is an indispensable parameter of the CASA model. However, the sparse distribution of the $R_s$ observation stations (five in the study area) resulted in it being difficult to interpolate accurate raster radiation data [73]. Therefore, the improved Yang hybrid model (YHM) was employed to calculate the $R_s$ of the 29 climatological stations in Yunnan province [73], using the daily climatological records. The model considers the attenuation from each atmospheric component when solar radiation passes through the atmosphere. The model can be described as follows:

$$R_s = \tau_c \int_{\Delta t} (\tau_{b,clear} + \tau_{d,clear}) I_0 \mathrm{d}t, \tag{10}$$

where $I_0$ is the solar irradiance at the top of atmosphere; $\Delta t$ is the time period of the calculated solar radiation; $\tau_{b,clear}$ and $\tau_{d,clear}$ are the solar beam radiative transmittance and the solar diffuse radiative transmittance under clear skies, respectively, which were determined by the transmittances of ozone, water vapor, gas mixture, the Rayleigh effect, and aerosols in the atmosphere; and $\tau_c$ is the radiative transmittance of cloud, which is a linear function of sunshine duration. The improved estimation of parameter $\tau_c$ was undertaken according to the method proposed by Wang et al. [74].

## 4. Results and discussion

In this section, the reliability of estimated NPP was firstly verified, as well as the accuracy of NDVI fusion and total solar radiation calculation. Then, the spatial and temporal variation of NPP in Yunnan was characterized. In order to unveil the climatic impacts on NPP, their correlations were analyzed at an annual and a monthly scale, respectively. The lagged effects of precipitation were also considered. Finally, the dominating climate factor for NPP variation was found for different time stages in the past three decades.

### 4.1 Results validation

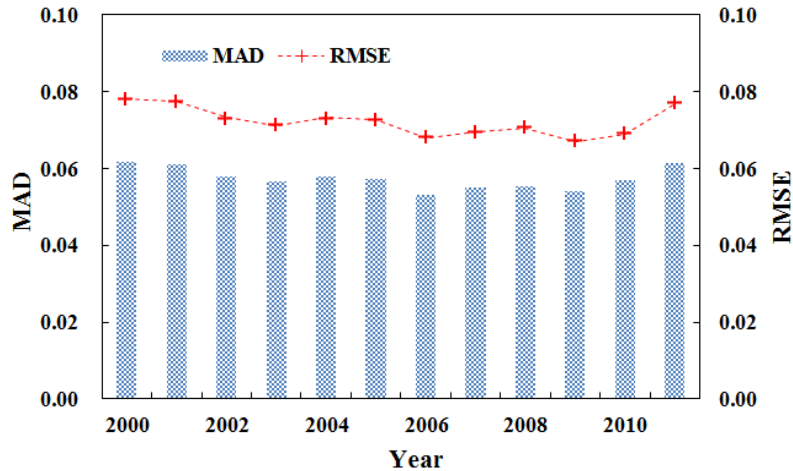

**Figure 4.** Inter-annual variation of the mean root-mean-square error (RMSE) and mean absolute difference (MAD) for the fusion method.

### 4.1.1. Simulated validation of the multi-sensor fusion

In order to validate the feasibility of the multi-sensor fusion process in long-term NDVI prediction, simulated experiments were conducted based on the MODIS and GIMMS3g datasets for the mutual time period of 2000 to 2012. For each month, the fine-resolution NDVI ($NDVI_{\text{fusion}}$) from 2000 to 2011 was simulated with reference to the MODIS and MODIS-like images for the corresponding month in the year 2012. The predicted fusion data ($NDVI_{\text{fusion}}$) were then validated with the true MODIS NDVI, both qualitatively and quantitatively. The statistics of root-mean-square error (RMSE) and mean absolute difference (MAD) for each year were computed and are shown in Figure 4. The fused NDVI presents an absolute error of less than 0.06 compared with the original MODIS data, and the RMSE is around 0.07. Most importantly, the fine fusion results maintain a stable accuracy as the years between the prediction date and reference date increase. It is therefore reasonable to predict the fine-resolution monthly NDVI before the year of 2000 using the multi-sensor fusion method. What is more, a qualitative comparison of the results for April 2000 is shown in Figure 5 as an example, where it can be observed that the spatial resolution of the predicted NDVI has been clearly improved. The spatial distribution of the data is highly consistent, and only slight differences can be found. As shown in the two green rectangles in Figure 5 (e) and Figure 5 (f), although some regions with low NDVI have been overestimated in the fusion result, the vegetated land area is well predicted. It means that the fusion results are useable in the vegetation related studies.

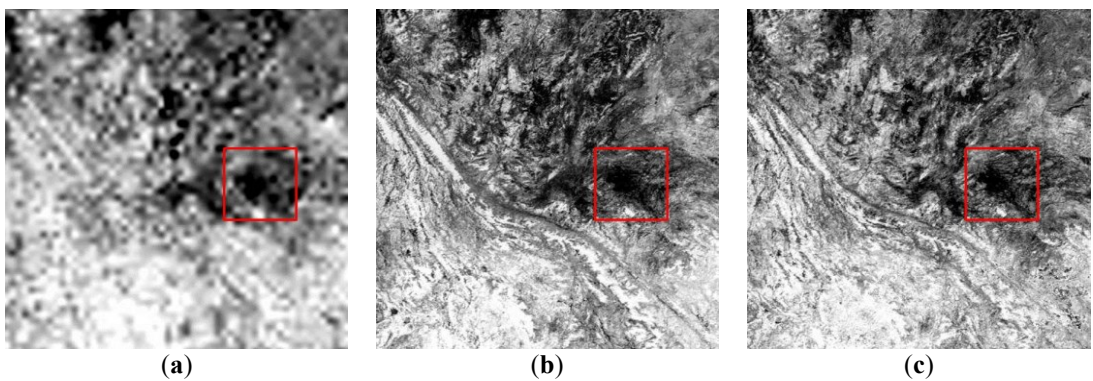

(**a**)　　　　　　　　(**b**)　　　　　　　　(**c**)

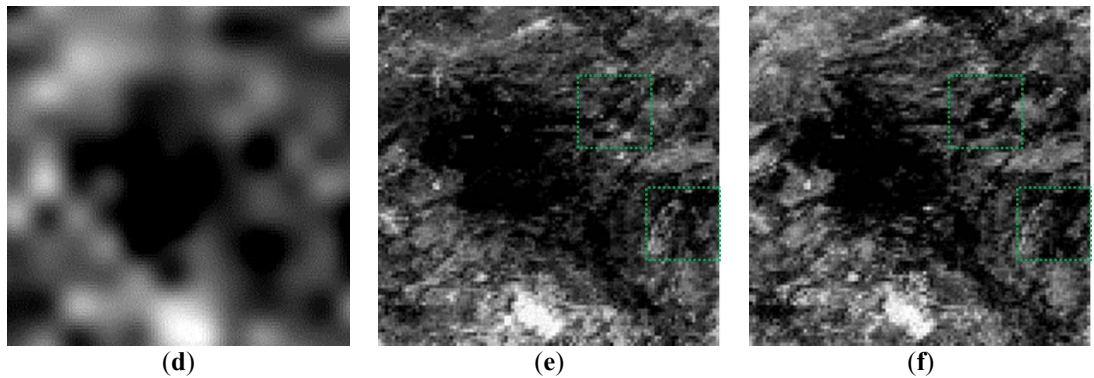

(**d**)   (**e**)   (**f**)

**Figure 5.** Qualitative comparison of the fusion method for April 2000: (**a**) resampled result of the original MODIS-like NDVI; (**b**) fusion result $NDVI_{\text{fusion}}$; (**c**) real MODIS data; (**d**), (**e**), (**f**) are the enlarged display for the part in the red rectangle of (a), (b), (c), respectively.

### 4.1.2. Cross-validation of the total solar radiation calculation

Cross-validation experiments were applied to examine the accuracy of the YHM model based on the quality-controlled monthly solar radiation measurements ($R_{\text{station}}$) [75], of five stations in Yunnan (station numbers 56651, 56739, 56778, 56959, and 56985, respectively). The records of four stations were used to obtain parameter $\tau_c$ and the model solar radiation ($R_{\text{model}}$), and then the $R_{\text{station}}$ records of the remaining one station (test station) were applied to validate the calculated $R_{\text{model}}$ of the corresponding site. The experiment was undertaken five times until every station was tested. The statistics of the correlation coefficient ($r$), RMSE, and mean absolute relative difference (MARD, the mean absolute value of the ratio between the error and true data) are listed in Table 2. The results show that the modeled solar radiation has a good consistency with the observed records, with a high $r$ value of much greater than 0.83 and a low MARD of around 8%.

**Table 2.** Statistics for the cross-validation of the improved YHM model.

| Test station no. | $r$ | RMSE (MJ m$^{-2}$) | MARD (%) |
|:---:|:---:|:---:|:---:|
| 56651 | 0.88 | 48.24 | 7.31 |
| 56739 | 0.84 | 52.76 | 8.71 |
| 56778 | 0.94 | 48.60 | 7.67 |
| 56959 | 0.85 | 46.86 | 7.96 |
| 56985 | 0.83 | 53.72 | 8.41 |

### 4.1.3. Validation of the estimated NPP

The reliability of the obtained NPP was verified using the field measurements from Luo's study (Luo 1996). In total, 59 plots were picked through matching the vegetation type with the land-cover map used in this study. In addition, in order to prove the necessity and advantage of the NDVI data processes, the NPP based on the original GIMMS3g NDVI and the MODIS-like NDVI were also calculated and validated. The verification of the three types of modeled NPP is shown in Figure 6, where it can be seen that the NPP calculated with the fused NDVI is the most accurate, with $r$ reaching 0.79 ($p$<0.001). The accuracy of the NPP obtained with the original GIMMS3g NDVI is the worst, with NPP obviously overestimated in almost all the plots ($r$=0.33). Meanwhile, the performance of the NPP based on the normalized MODIS-like NDVI is much better ($r$=0.44), due to the NDVI being normalized to be the same as the MODIS data, with a higher precision and lower values. Moreover, the spatial resolution of the final fused NDVI has been promoted, so that it reduces the overestimation and precision loss caused by the spatial heterogeneity when modeling NPP. In general, although the scale effect and representative errors might exist in the validation, both the necessity of the NDVI processes and the reliability of the modeled NPP have been effectually proven in the study area.

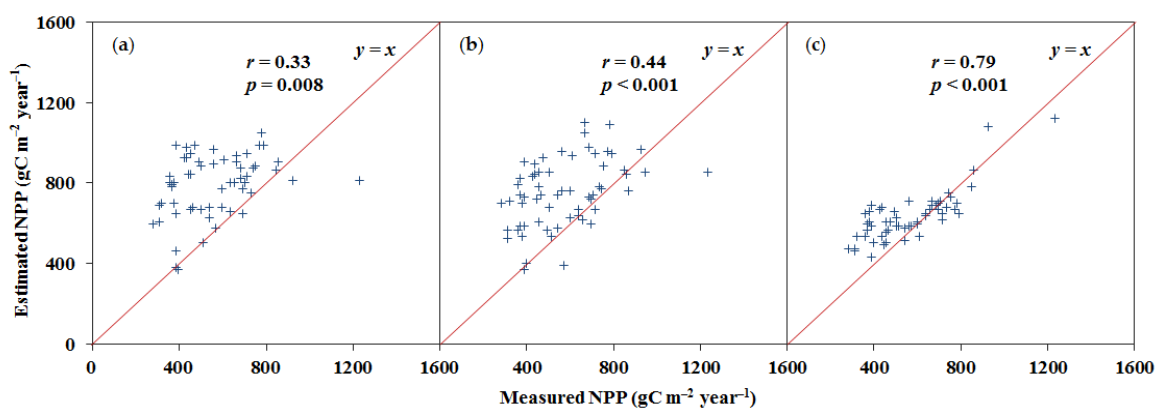

**Figure 6.** The relationships between estimated NPP and field-measured data ($N$=59). The three scatter plots are the NPP results based on (**a**) the original GIMMS3g NDVI; (**b**) the normalized MODIS-like NDVI; and (**c**) the final fused NDVI.

### 4.2. NPP spatial distribution and variation trends

#### 4.2.1. NPP spatial distribution

The spatial distribution of the mean annual NPP for the past 33 years is shown in Figure 7 (a). On the whole, the NPP in Yunnan province gradually decreases from the southwest to the northeast. The mean annual NPP is generally higher than 1000 gC m$^{-2}$ year$^{-1}$ in most of the southwest area, which is located close to the frontier, where the latitude and altitude are relatively low. The lack of human activities and the warm climate in this area benefit the growth of vegetation. However, the mean annual NPP is less than 500 gC m$^{-2}$ year$^{-1}$ in extensive regions of northwest Yunnan, which is a part of the Qinghai–Tibet Plateau, with elevations mostly higher than 4000 m. The cold and harsh climate in this area limits the vegetation growth, which is also the reason for the low NPP in the northeast area. There are also many regions with an NPP of less than 500 gC m$^{-2}$ year$^{-1}$ in central Yunnan, such as the provincial capital of the city of Kunming, as a result of urban construction and expansion. In general, most of Yunnan shows a relatively high NPP exceeding 800 gC m$^{-2}$ year$^{-1}$, but some regions have lower NPP values due to the harsh climate or human activities.

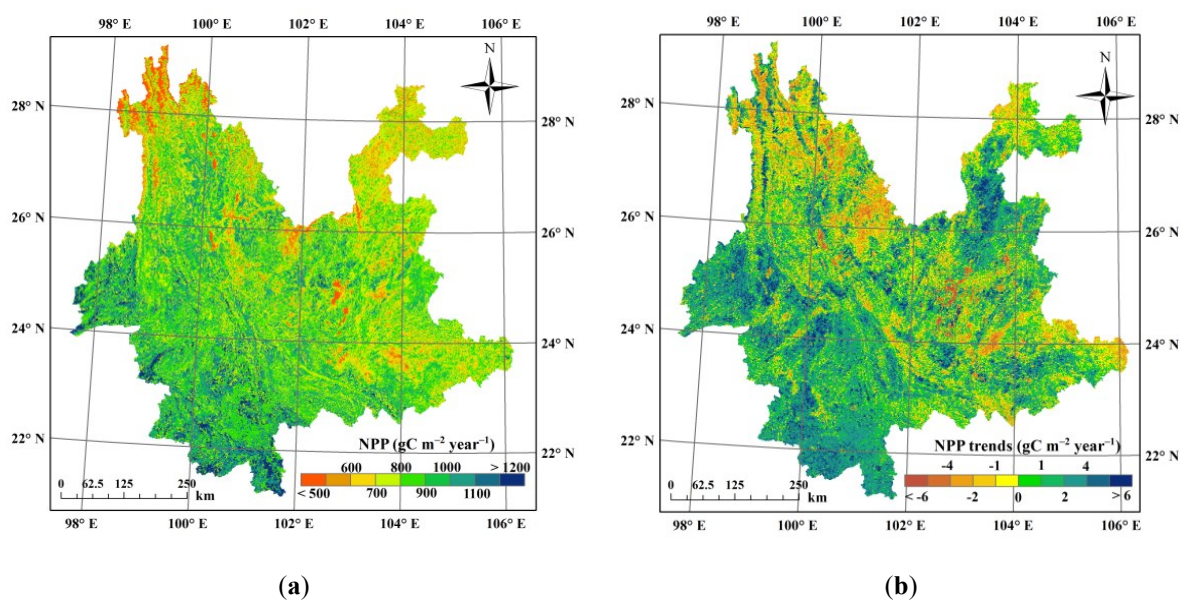

**Figure 7.** The spatial patterns of: (**a**) mean annual NPP over the past 33 years; and (**b**) annual NPP trends (the slope of NPP inter-annual variation).

#### 4.2.2. Annual NPP variation

The inter-annual changes of the mean annual NPP in Yunnan province are shown in Figure 8 (a). Overall, the mean annual NPP showed fluctuating growth from 1982 to 2014, with a total increasing trend of 0.98 gC m$^{-2}$ per year ($r$=0.38, $p$<0.05). However, the inter-annual variation of NPP was not consistent over the entire study period, but showed three distinct stages. It can be clearly observed that the NPP experienced a decreasing trend from 1982 to 1992 (slope= −3.04 gC m$^{-2}$ year$^{-1}$); it then sharply increased at a rate of 5.70 gC m$^{-2}$ year$^{-1}$ until 2002; and finally slightly decreased again between 2002 and 2014 (slope= −2.22 gC m$^{-2}$ year$^{-1}$). Therefore, the increasing trend of the NPP over the study period was mainly due to the increment from 1992 to 2002, because the NPP in the other two stages presented completely opposite variation trends.

The spatial pattern of the annual NPP trends (the slope of the NPP inter-annual variation) in the past 33 years is shown in Figure 7 (b). The annual NPP increased in 68.07% of the study area, which is more than twice the region with decreased NPP. In particular, 37.90% of the area showed a significantly increased NPP trend ($r$>0, $p$<0.05), while the rate for the significantly decreased area was only 9.61% ($r$<0, $p$<0.05). Almost all of southwest Yunnan showed increasing trends, with values higher than 2 gC m$^{-2}$ per year. This indicates that the NPP in southwest Yunnan is not only high, but has also presented a significant increasing trend over the past 33 years. Obvious increasing trends also occurred in large regions in the northeast, which shows a relatively low NPP. Meanwhile, for the high-altitude district in the northwest, although some regions showed an increasing trend, the annual NPP for most of the area presented obvious decreasing trends. Urban expansion also led to the apparent decline of NPP for some regions in central Yunnan.

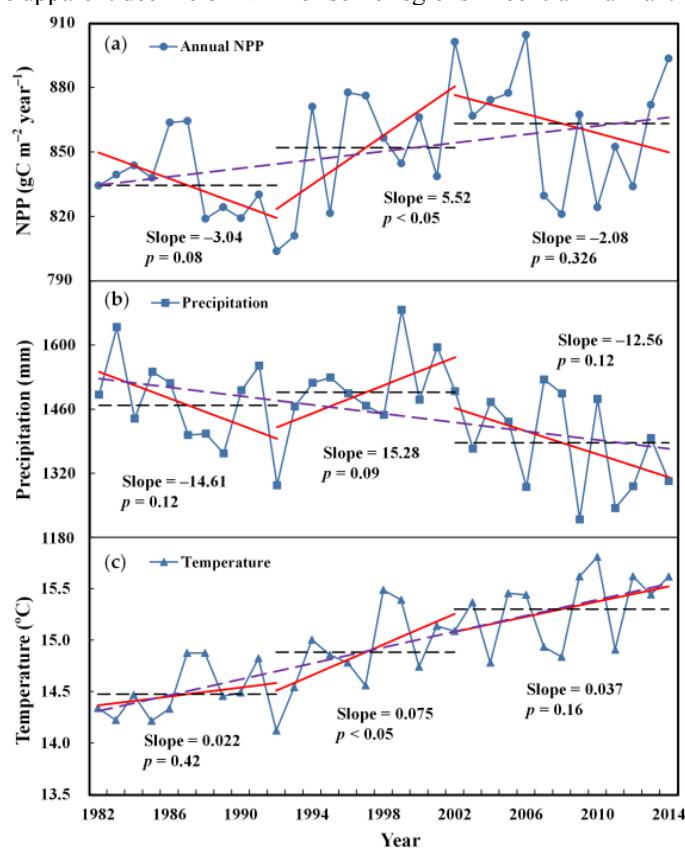

**Figure 8.** Inter-annual variation of NPP and climatic factors in Yunnan province from 1982 to 2014: (**a**) mean annual NPP; (**b**) annual accumulated precipitation; and (**c**) annual mean temperature. The dotted purple lines are the overall linear regress for the 33 years, and the solid red lines show the linear regress for each stage.

*4.3. Relationship between NPP and climate*

4.3.1. Correlations between NPP and climatic factors at an annual scale

The interaction between NPP and climatic factors is one of the most important issues for carbon cycle research. The inter-annual changes of mean temperature and accumulated precipitation for the study area are shown in Figure 8 (b) and (c). It can be seen that the temperature steadily increased in the three stages, with an overall increasing trend of 0.039°C per year ($r$=0.81, $p$<0.01), and the total increment reached 1.28°C from 1982 to 2014. The benefit of continuous climate warming was responsible for the overall growth in NPP (Nemani et al. 2003). Temperature also showed a significant positive correlation with NPP at an annual scale, with $r$=0.41 ($p$<0.05). Precipitation presented three completely parallel stages to NPP, with trends of −14.61 mm year$^{-1}$, 15.28 mm year$^{-1}$, and −12.56 mm year$^{-1}$, respectively. The relative magnitudes of the variation trends for the three stages also agreed with the NPP. However, the precipitation showed an overall variation trend that was the opposite to NPP, with a decreasing trend of −4.79 mm per year ($r$= −0.43, $p$<0.05). An abnormally negative correlation was observed between annual NPP and precipitation, in spite of the fact that the relationship was not significant ($r$= −0.22, $p$>0.1). This indicates that increased precipitation led to reduced NPP for vegetation in Yunnan province.

### 4.3.2. Correlations between NPP and climatic factors at a monthly scale

In order to conduct further studies of the relationships between NPP and climatic factors, the intra-annual variations of monthly NPP, climatic factors, and their correlations are shown in Figure 9, as well as the important parameter of total solar radiation. It can be seen that there was an obviously higher NPP in summer and low accumulation in winter, with the monthly NPP being more than 90 gC m$^{-2}$ per month from May to September, and only about 30 gC m$^{-2}$ month$^{-1}$ in the three months of winter. This characteristic of NPP is the result of the similar uneven distribution of precipitation and temperature at a monthly scale. The mean temperature in summer was more than the double that in winter, with a difference of more than 10.46°C. The heterogeneous intra-annual distribution of precipitation was even greater. The half-year from May to October featured 82.88% of the annual precipitation. The suitable temperature and abundant precipitation provide good conditions for vegetation growth from May to October, while the cold and dry climate suppresses photosynthesis and leads to low NPP in the other months. The period from May to October is the "growing season", with high NPP and a superior hydrothermal environment, and the other six months can be called the "dry season", with low NPP and a harsh climate.

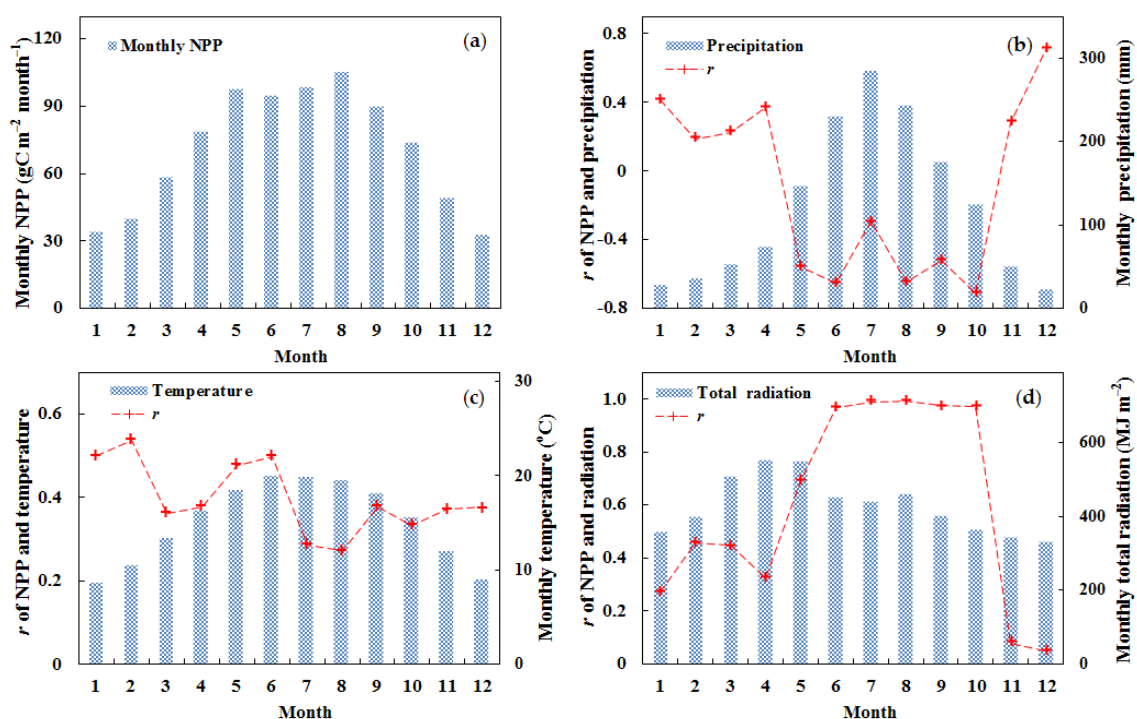

**Figure 9.** The intra-annual variation of: (**a**) monthly NPP; (**b**) monthly precipitation and its correlation with NPP; (**c**) monthly temperature and its correlation with NPP; and (**d**) monthly total solar radiation and its correlation with NPP.

Figure 9 also shows the difference in the correlation coefficients ($r$) between monthly NPP and the climatic variables for each month. Temperature presents a relatively stable positive correlation with NPP throughout the year, with a relatively weaker impact in summer when the warm climate is at the optimum level for photosynthesis. Nevertheless, highly fluctuating correlation coefficients between NPP and precipitation and total solar radiation are observed in the intra-annual variation, and completely antipodal distributions are found between them. Total solar radiation fully controls the change of NPP from June to October, with amazingly high correlation coefficients of greater than 0.97 ($p<0.01$). The relationship in May is also significant ($r=0.69$, $p<0.01$). In contrast, precipitation shows a significant negative impact on the vegetation growth in the six months of the growing season. This phenomenon is the reason for the negative correlation between precipitation and NPP at an annual scale. The significant negative impacts during the period lead to the weak negative correlation between NPP and precipitation at an annual scale because NPP in the growing season amounts to 65.7% of the annual amount. This greatly affects the variation of annual NPP.

The negative impact of precipitation on NPP in the growing season is the result of the warm temperature and abundant precipitation during this period, whose mean values are 18.61°C and 200.31 mm per month. The least rainfall in the growing season occurred in the year 2011, but the average amount was still as high as 67.09 mm per month. In this condition, total solar radiation, as the energy source of photosynthesis, certainly dominates the growth of vegetation, with a very close correlation to NPP. However, due to the rainfall being abundant for photosynthesis during this period, the increased precipitation does not benefit the vegetation and actually suppresses its growth. This is because the increased precipitation means more overcast skies, which prevents the transfer of sunlight and lessens the total amount of solar radiation arriving at the vegetation canopy. After verification, the precipitation and total solar radiation show a significant negative mean correlation during the period, with $r=-0.64$ ($p<0.01$). Therefore, the increase of precipitation in the growing season does not promote the growth of vegetation, but instead weakens the photosynthesis by cutting down the energy source from solar radiation.

On the other hand, precipitation shows a positive influence on the variation of NPP in the time of the dry season. This is the result of the extreme lack of water for vegetation growth in these months, when the mean precipitation is 42.58 mm per month, less than one-fifth the level of the growing season. Thus, precipitation replaces radiation as the main limitation for photosynthesis, as increased precipitation promotes vegetation growth by providing the necessary element of water.

### 4.3.3. Lagged impact of precipitation on NPP

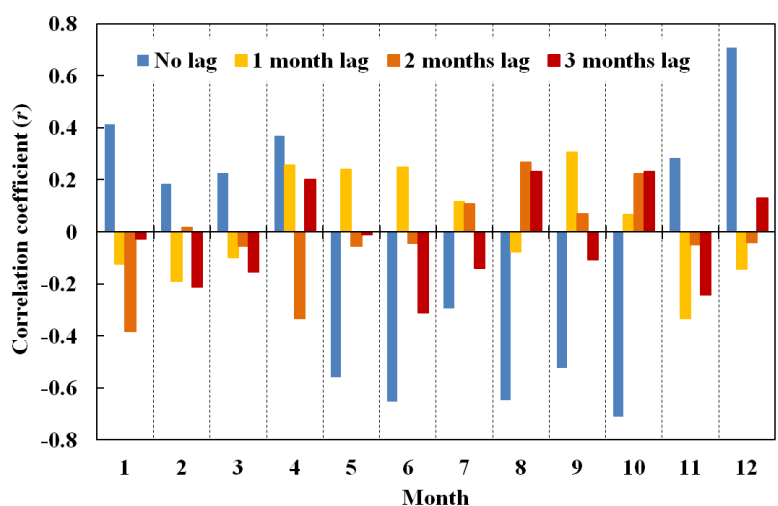

**Figure 10.** Lagged correlations between monthly NPP and precipitation for different months.

The delayed effect of precipitation on NPP should also be considered, and time lags of 0–3 months were considered according to previous studies [76]. From January to December, the correlations between monthly NPP and precipitation in the same month or 1–3 months previous were calculated and compared, as shown in Figure 10. Although NPP in the growing season shows a significant negative correlation with precipitation, optimal positive correlations (meaning the highest correlation coefficients) can be observed with the precipitation of one or two months before. Except for August, during which the optimal time lag is two months, the NPP of the other months in the growing season all show one month lag with precipitation. This indicates that increased precipitation in the previous one or two months can promote the growth of vegetation, because soil can store water from rainfall and provide water to the vegetation in the following months. Meanwhile, this phenomenon cannot be found for the dry season months, during which the optimal correlation between NPP and precipitation appears in the identical month, without lag. This might be the result of the urgent demand for water in these dry months, so vegetation immediately responds to the precipitation, without time lag. This indicates that precipitation has an important impact on the variation of NPP in the study area, even though a negative correlation is observed in the growing season.

### 4.4. Driving forces for the inter-annual variation of NPP in the three stages

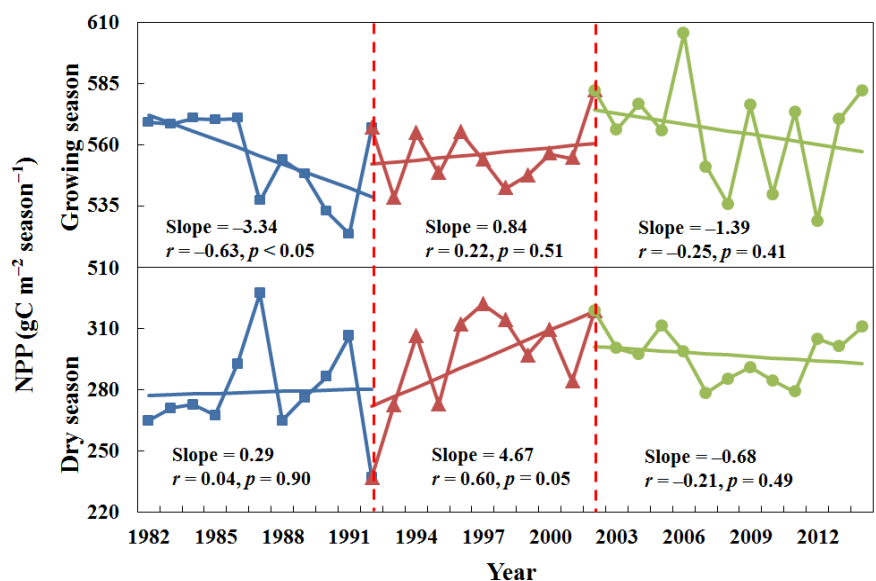

**Figure 11.** Inter-annual variation of NPP in the growing and dry seasons for the three stages.

Although precipitation presented three completely parallel stages to NPP, it was not enough to explain the three-stage inter-annual variation of NPP. The variation of NPP in the growing season and dry season was therefore calculated to further study the driving forces for NPP change in the three stages, as shown in Figure 11. It can be seen that the changes of annual NPP in the three periods were dominated by the amounts in the different seasons. For the stage from 1982 to 1992, the decline of annual NPP was apparently dominated by the NPP change in the growing season (slope= −3.34), and the NPP increment from 1992 to 2002 was caused by the NPP variation in the dry season (slope= 4.67). Meanwhile, the decrease of NPP from 2002 to 2014 was the result of the decreased NPP in both the growing season (slope= −1.39) and the dry season (slope= −0.68). The variation trends of the climatic factors in the related seasons and stages were calculated to find the reasons for the NPP change, as shown in Table 3. For the time periods of the growing season from 1982 to 1992 and the two seasons from 2002 to 2014, precipitation showed the same decreasing variation trends as NPP (−8.25, −9.83, −7.43 mm per year, respectively). Precipitation did not show an obvious decreasing trend in the dry season from 1992 to 2002 (only −1.52 mm per year) when the NPP increased. Instead, temperature presented a very significant growth of 0.15°C per year in this period, which was much greater than the temperature rise in the other stages.

**Table 3.** Variation trends of NPP and climatic factors in the related seasons and stages.

| Time period | Season | NPP (gC m$^{-2}$ year$^{-1}$) | Precipitation (mm year$^{-1}$) | Temperature (°C year$^{-1}$) |
|---|---|---|---|---|
| 1982–1992 | Growing season | −3.34 | −8.25 | −0.014 |
| 1992–2002 | Dry season | 4.67 | −1.52 | 0.15 |
| 2002–2014 | Growing season | −1.39 | −9.83 | 0.054 |
| | Dry season | −0.68 | −7.43 | 0.021 |

The variation trends of NPP and climatic factors at the pixel level were also calculated and analyzed for the four groups of data. For the area where NPP showed the same variation trends as the mean seasonal NPP, statistics of the climate variation trends were calculated and are shown in Table 4. Taking the example of the growing season from 1982 to 1992, the percentage of area with increased or decreased precipitation/temperature was counted among the pixels with decreased NPP. For the growing season of the period from 1982 and 1992, 95.54% of the study area showed a decreased NPP in total. Among these regions, 78.63% showed decreased precipitation, with a mean trend of −12 mm per year. The continuous reduction in precipitation was likely responsible for the reduction of NPP, because it would cause persistent droughts. The growth of vegetation was also affected by the continuous decline of temperature in 72.66% of the area, due to the decreased biological activity. The decrease of precipitation would also be the reason for the decline of NPP from 2002 to 2014. Especially in the dry season, 99.07% of the area with a decreased NPP presented a precipitation decrease, with a mean trend of −7.80 mm season⁻¹. The value was 19.21% of the monthly precipitation amount in that period. The percentage of area and mean trend in the growing season also reached 63.41% and −12.22 mm per year. The significant precipitation decreases in each season have caused frequent droughts during this period, which were the reason for the NPP decline, most notably the four-year extreme drought from 2009 to 2012. Meanwhile, for the dry season from 1992 to 2002, although the mean seasonal precipitation slightly decreased, more rainfall occurred in 41.08% of the area. As a result, the rapidly increasing temperature trend of 0.15°C per year greatly promoted the vegetation growth. The climate warming benefited the vegetation even more in the dry season, when the temperature was relatively low, with a mean value of only 11.03°C. In general, it can be concluded that the increment of NPP from 1992 to 2002 was mainly caused by the growth of the dry season due to the significant climate warming, and the decline in NPP from 1982 to 1992 and from 2002 to 2014 was due to the frequent droughts caused by the precipitation decrease during these periods.

**Table 4.** The percentage of decreased and increased pixels of precipitation and temperature within the area showing the same variation trends as the seasonal NPP.

| Time period | Season | NPP trend | Decreased pixels (%) | | Increased pixels (%) | |
|---|---|---|---|---|---|---|
| | | | $P$ | $T$ | $P$ | $T$ |
| **1982–1992** | Growing | Decreasing | 78.63 | 72.66 | 21.37 | 27.34 |
| **1992–2002** | Dry | Increasing | 58.92 | 0 | 41.08 | 100 |
| **2002–2014** | Growing | Decreasing | 63.41 | 5.95 | 36.59 | 94.05 |
| | Dry | Decreasing | 99.07 | 26.98 | 0.93 | 73.02 |

Note: $P$ denotes the precipitation; $T$ denotes the temperature.

## 5. Conclusion

In this paper, we have conducted a monitoring study of the regional carbon cycle in Yunnan province of southwest China, from 1982 to 2014, at a 1-km scale. Based on the CASA model and multi-source data, a feasible framework for long-term and accurate NPP was presented: NDVI time series were obtained through fusing the GIMMS3g data and MODIS data to combine their respective advantages, and simulated experiments confirmed the reliability and stability of the fusion results, with a stable MAD of less than 0.06; $R_s$ was calculated using the improved YHM model, and cross-validation showed that the MARD was around 8% and $r$ was much higher than 0.8. The final estimated NPP results presented good consistency with the field measurements ($r$=0.79), and huge accuracy improvements were observed when compared to the NPP based on the original NDVI.

Based on the simulated NPP time series, spatial and temporal analyses were undertaken and some interesting conclusions were found. In spatial terms, 68.07% of the study area showed an increasing NPP trend, and the NPP in southwest Yunnan was both higher and increased more rapidly overall. Annual NPP variation showed three distinct stages in the past 33 years, and the overall increasing trend was dominated by the growth during 1992–2002. Precipitation presented the same three stages as NPP, but the non-significant negative correlations at an annual scale cannot explain its impact on NPP, which was the result of the enormously uneven intra-annual distribution of NPP and climatic factors. There was a significantly negative correlation in the growing season, due to the increased precipitation reducing the total radiation. Meanwhile, one or two months lagged positive correlations were observed in the growing season, and no lag was observed in the dry season, which proves the important impact of precipitation on NPP. Furthermore, the frequent droughts caused by the precipitation decrease led to the NPP decline during 1982–1992 and 2002–2014; NPP rapidly increased from 1992 to 2002 due to the significant climate warming, when the precipitation varied only slightly.

In summary, this study has proposed an efficient framework that can be applied to study the regional carbon cycle at a long-term and accurate level, which will be beneficial for the systematic understanding of the global carbon cycle. However, further efforts should be made to investigate the more detailed spatial and temporal patterns of NPP in such complex terrain, such as the specific variation trends and the response of NPP to climate in different elevation zones. Furthermore, the relationships between NPP and more environmental factors (i.e. other climatic factors, soil parameters, atmospheric condition, land-use/land-cover changes, disaster disturbance, human activities, and so on) also need to be investigated, in order to obtain a better understanding of the influencing mechanism of the regional carbon cycle.

**Acknowledgments:** This research was supported by the National Natural Science Foundation of China (41422108, 41661134015), Cross-disciplinary Collaborative Teams Program for Science, Technology and Innovation of the Chinese Academy of Sciences, and the Open Research Fund of Key Laboratory of Digital Earth Science, Institute of Remote Sensing and Digital Earth, Chinese Academy of Sciences (under Grant No. 2016LDE004). We would like to thank the China Meteorological Administration for providing the radiation and meteorological data. We would also like to thank the NDVI data providers of the NASA Ames Ecological Forecasting Lab and Earth Observing System. Special thanks are given to all the people who have provided helpful comments and suggestions.